\documentstyle[aps,prb,epsfig,multicol]{revtex}
\begin{document}
\title{The effect of temperature jumps during polymer crystallization}
\author{Jonathan P.~K.~Doye$^{1,2}$ and Daan Frenkel$^2$}
\address{$^1$ University Chemical Laboratory, Lensfield Road, Cambridge CB2 1EW, UK} 
\address{$^2$ FOM Institute for Atomic and Molecular Physics, Kruislaan 407,\\ 
1098 SJ Amsterdam, The Netherlands}
\date{\today}
\maketitle
\begin{abstract}
Temperature changes during the growth of lamellar polymer crystals give rise 
to steps on the surface of the crystals.
It has recently been suggested that these steps could provide important insights into the 
mechanism of polymer crystallization. 
In particular, a characterization of the profiles of these steps
might reveal the fixed-point attractor that underlies a recently proposed 
crystallization mechanism.
Here we examine this hypothesis by performing simulations of such temperature jumps
using the Sadler-Gilmer model. We find that for this model the 
step profiles do reveal the fixed-point attractor. However, 
for temperature decreases they also reflect the rounding of the crystal edge that occurs in this
model and for temperature increases they also reflect the fluctuations in the thickness present in the crystal.
We discuss the implications of these results for the interpretation of experimental step profiles.
\end{abstract}
\pacs{}

\begin{multicols}{2}
\section{Introduction}

\begin{figure}
\begin{center}
\epsfig{figure=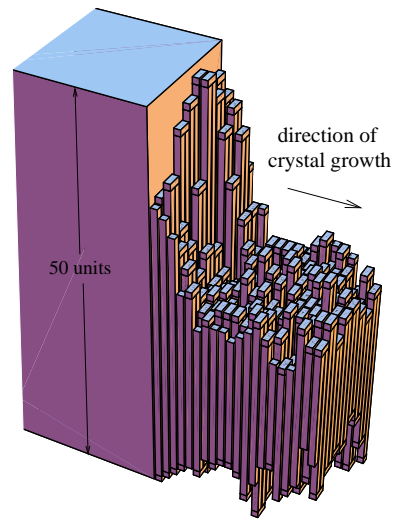,width=7.3cm}
\vglue -0.2cm
\begin{minipage}{8.5cm}
\caption{\label{fig:3D}Cut through a polymer crystal which was produced by the growth of 
twenty successive layers on a surface with a uniform thickness of 50 units using 
the model described in Refs. \protect\onlinecite{Doye98b} and \protect\onlinecite{Doye98d}.
Folds occur at the top and bottom surfaces, and 
the stems are represented by vertical cuboids. The cut is 16 stems wide. 
}
\end{minipage}
\end{center}
\end{figure}

Upon crystallization from solution and the melt many polymers form lamellae
where the polymer chain traverses the thin dimension of the crystal many times folding back
on itself at each surface.\cite{Keller57a,Keller96a}
(The crystal geometry is shown by the example configuration in Figure \ref{fig:3D})
Although lamellar crystals were first observed over forty years ago
their physical origin is still controversial.
In particular the explanations for the dependence of the lamellar thickness on temperature
offered by the two dominant theoretical approaches---the Lauritzen-Hoffman surface nucleation
theory\cite{Lauritzen60,Hoffman76a,Hoffman97} and the entropic barrier 
model of Sadler and Gilmer\cite{Sadler84a,Sadler86a,Sadler87d,Sadler88a}---differ greatly.\cite{theorynoteb}

One of the common features of the two theories is that they both argue that the observed crystal thickness
is close to the thickness at which the crystal growth rate is a maximum. 
However, recently a new description of the mechanism of thickness selection has been presented.\cite{Doye98b,Doye98d,Doye98e}
In this approach the observed thickness corresponds instead to the one thickness, $l^{**}$, at which
growth with constant thickness can occur. Crystals initially thicker (thinner) than $l^{**}$
will thin (thicken) as the crystals grow until the thickness $l^{**}$ is reached.
This dynamical convergence can be described by a fixed-point attractor which relates the thickness of a layer
to the thickness of the previous layer. 
The value of the thickness at the fixed point is $l^{**}$.

This mechanism has been found for two simple models of polymer crystallization.\cite{Doye98b,Doye98d,Doye98e} 
In the first model the polymer crystal grows, as in the Lauritzen-Hoffman theory, by the successive deposition of 
stems (a stem is a straight portion of the polymer chain that traverses the thin dimension of the lamella) 
across the growth face.\cite{Doye98b,Doye98d}
A configuration produced by this model is shown in Figure \ref{fig:3D} to illustrate
the mechanism.
The crystal thins down from the initial thickness to $l^{**}$ within five to ten layers
and then continues to grow at that thickness
The second model is that used by Sadler and Gilmer, the behaviour of which 
they interpreted in terms of an entropic barrier. 
In this Sadler-Gilmer (SG) model the connectivity of the polymer is modelled implicitly, 
the growth face can be rough, and lateral correlations along the growth face can be weak.
That we find the same mechanism in these two very different models is 
a sign of its generality.

\vglue -0.3cm

Further support for the new mechanism is provided by the experimental observation that a temperature
change during crystallization produces a step on a lamella.\cite{Bassett62,Dosiere86a} 
This step is a result of the thickness of 
the crystal dynamically converging to $l^{**}$ for the new temperature as the crystal grows.
Furthermore it has been suggested that a detailed characterization of the step profiles by 
atomic-force microscopy could 
allow the fixed-point attractors that underlie the mechanism to be obtained.\cite{Doye98b,Doye98d,Doye98e}
In this paper we examine this suggestion more carefully by performing simulations of
temperature jumps for the Sadler-Gilmer model.\cite{GGWstep} 
In particular, we investigate the effect that rounding of the crystal profile near to the growth face 
and fluctuations in the crystal thickness may have on the shape of the steps.
It is hoped that this work will aid the experimental interpretation of step profiles.
 
\begin{center}
\begin{figure}
\epsfig{figure=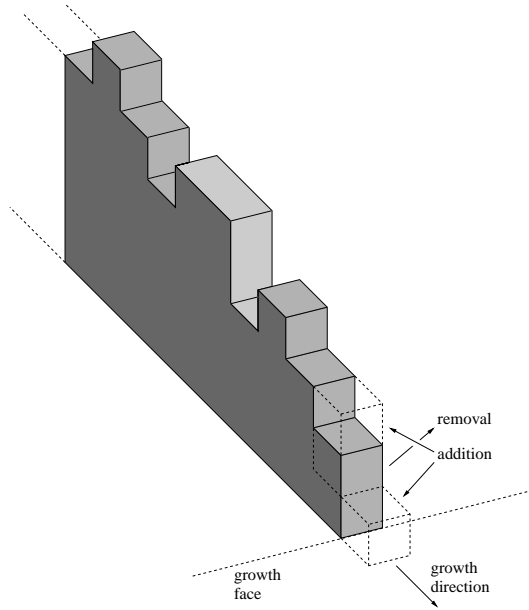,width=7.0cm}
\vglue 1mm
\begin{minipage}{8.5cm}
\caption{\label{fig:moves} A schematic picture of a two-dimensional slice (perpendicular to the
growth face) through a lamellar polymer crystal which forms the basis of the two-dimensional
version of the Sadler-Gilmer model. 
The three possible changes in configuration allowed by the model are shown (the dashed lines
represent the outline of the possible new configurations).
}
\end{minipage}
\end{figure}
\end{center}

\section{Methods}

In the SG model the growth of a polymer crystal 
results from the attachment and detachment of polymer units at the growth face. 
The rules that govern the sites at which these processes can occur 
are designed to mimic the effects of the chain connectivity.
In the original three-dimensional version of the model,
kinetic Monte Carlo simulations were performed to obtain many realizations of the polymer
crystals that result. 
Averages were then taken over these configurations to get the properties of the model.\cite{Sadler84a}
Under many conditions the growth face is rough and the correlations between stems in
the direction parallel to the growth face are weak. 
Therefore, an even simpler two-dimensional version of the model was developed in which 
lateral correlations are neglected entirely, 
and only a slice through the polymer crystal perpendicular to the growth face is considered.\cite{Sadler86a,Sadler88a}
The behaviour of this new model was found to be very similar to the original three-dimensional
model.

The geometry of the model is shown in Figure \ref{fig:moves}.
Changes in configuration can only occur at the outermost stem and stems behind
the growth face are `pinned' because of the chain connectivity.
There are three ways that a polymer unit can be added to or removed from the crystal:
(1) The outermost stem can increase in length upwards. (2) A new stem can be initiated
at the base of the previous stem.
(3) A polymer unit can be removed from the top of the outermost stem.

The ratio of the rate constants for attachment ($k^+$) and detachment ($k^-$) of a polymer
unit are related to the thermodynamics of the model through 
\begin{equation}
k^+/k^-=\exp(-\Delta F/kT), 
\label{eq:kratio}
\end{equation}
where $\Delta F$ is the change in free energy on addition of 
a particular polymer unit.
The above equation only defines the relative rates and not how the 
the free energy change is apportioned between the 
forward and backward rate constants.
We follow Sadler and Gilmer and choose $k^+$ to be constant.
We use $1/k^+$ as our unit of time.

In the model the energy of interaction between two adjacent crystal units
is -$\epsilon$ and the change in entropy on melting of the crystal is given
by $\Delta S=\Delta H/T_m=2\epsilon/T_m$, where $T_m$ is the melting temperature 
(of an infinitely thick crystal) and 
$\Delta H$ is the change in enthalpy.
It is assumed that $\Delta S$ is independent of temperature.
Here, as with Sadler and Gilmer, we do not include any contribution from chain folds 
to the thermodynamics.

From the above considerations it follows that the rate constants for detachment
of polymer units are given by
\begin{eqnarray}
\label{eq:kratio2}
k^-(i,j)&=&k^+ \exp(2\epsilon/kT_m - 2\epsilon/kT) \quad\quad i\ne1, i\le j \\
k^-(i,j)&=&k^+ \exp(2\epsilon/kT_m - \epsilon/kT)  \ \quad\quad i=1, i>j,
\end{eqnarray}
where $i$ is the length of the outermost stem and $j$ the length of the stem in the previous layer.
The first term in the exponents is due to the gain in entropy as a result of the removal of a unit 
from the crystal, and the second term is due to the loss of contacts between the 
removed unit and the rest of the crystal.

There are two ways to examine the behaviour of the model. 
In one approach the model is formulated in terms of a set of rate equations which can easily be solved numerically
to yield the steady-state solution of the model.\cite{Sadler86a,Sadler88a}
This is the method that we used for the most part in our previous study of the Sadler-Gilmer model.\cite{Doye98e}
However, as we wish to examine the evolution of the system towards the steady state, 
we use kinetic Monte Carlo to grow a set of representative crystals.\cite{GGWrate} 
In this work we deliberately start growing these crystals from a well-defined non-steady-state initial 
configuration---either a crystal of constant thickness different from $l^{**}$, or
a crystal grown at a different temperature.
Averages are then taken over these crystals to obtain information about the convergence of the 
system towards the steady state.

At each step in the kinetic Monte Carlo simulation a state, $b$, is randomly chosen from the three states
connected to the current state, $a$, with a probability given by
\begin{equation}
P_{ab}={k_{ab}\over\sum_{b'}k_{ab'}},
\end{equation}
and update the time by an increment
\begin{equation}
\Delta t= -{\log\rho\over\sum_{b}k_{ab}},
\end{equation}
where $\rho$ is a random number in the range [0,1].
Depending on the conditions we use from tens of thousands to millions of steps to grow each individual 
crystal, and then take averages over many thousands of crystals.

The version of the SG model used here has two variables: $kT_m/\epsilon$ and $T/T_m$. 
Here, as in our previous work on the SG model,\cite{Doye98e} we use $kT_m/\epsilon=0.5$, 
unless otherwise stated. Sadler and Gilmer have shown that the basic properties
of the model were independent of the value of $kT_m/\epsilon$ within 
the parameter range that they studied.\cite{Sadler88a}

\section{Results}

Before considering simulations of actual temperature jumps we examine a slightly simpler 
case, namely the convergence of the crystal thickness to $l^{**}$ when the initial configuration
is a crystal of constant thickness different from $l^{**}$.
These cases will provide a useful comparison to the convergence to a new $l^{**}$ caused
by a change in temperature.

\begin{center}
\begin{figure}
\epsfig{figure=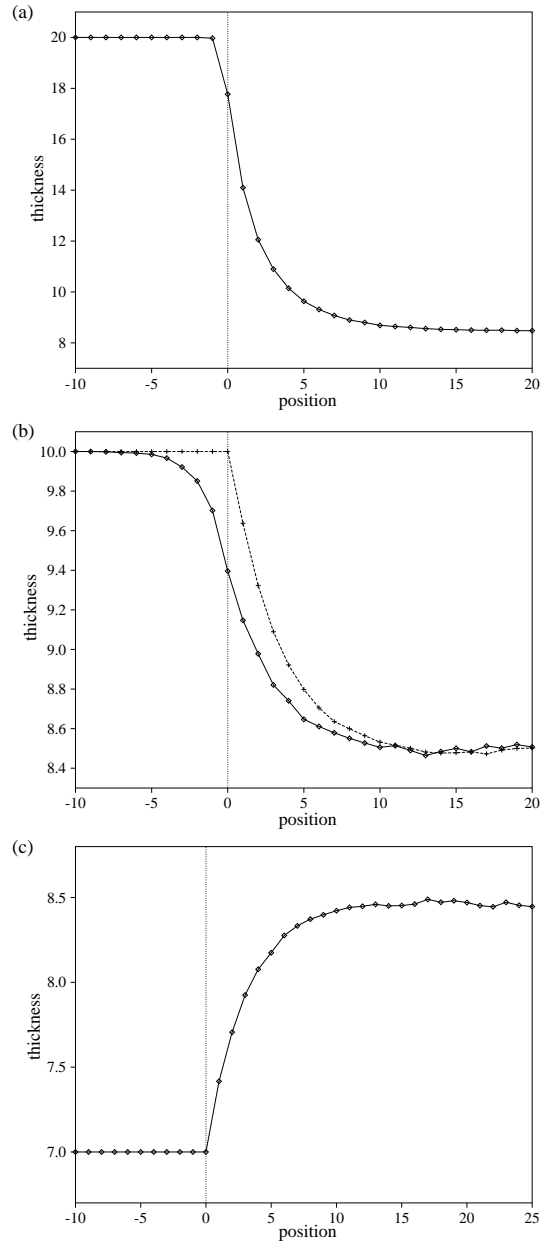,width=7.4cm}
\vglue 1mm
\begin{minipage}{8.5cm}
\caption{\label{fig:step.square} 
Average crystal profiles that result from the growth of an initial crystal that has 
a constant thickness of (a) 20, (b) 10 and (c) 7 polymer units.
For (a) and the solid line in (b) the zero of position is defined as the edge of the
initial crystal. For (c) and the dashed line in (b) the zero of position is defined as
the minimum position of the growth front during the simulation. $T=0.95T_m$.
}
\end{minipage}
\end{figure}
\end{center}

Figure \ref{fig:step.square} shows two example crystal profiles when the 
the thickness of the initial crystal is larger than $l^{**}$.
It can be clearly seen that the crystal quickly converges to $l^{**}$
whatever the initial thickness of the crystal leading to a downward step 
on the surface. 
For the crystal that is initially twenty units thick there is little backward
motion of the growth front in the early stages of the simulations because a thick crystal 
such as this is very stable. 
Therefore, after growth only the outer layer of the initial crystal has
a thickness different from its initial value. The step has a sharp downward edge.

By contrast, the step on the crystal that was initially ten units thick 
is more rounded (Figure \ref{fig:step.square}b). The outer layers of the initial crystal
are now noticeably less than ten up to about five layers back into the initial 
crystal and the curvature of the step changes from negative to positive around the
initial position of the edge of the crystal.

It is also worth noting that we previously found that the initial growth rate increases with the thickness
of the initial crystal.\cite{Doye98e} The cause of this behaviour is similar to that
for the more pronounced rounding of the steps that result from growth on thinner crystals.
Both are related to the smaller number of detachment steps in the kinetic Monte Carlo simulations
when the initial crystal is thicker because of the greater thermodynamic driving force for growth.

\begin{center}
\begin{figure}
\epsfig{figure=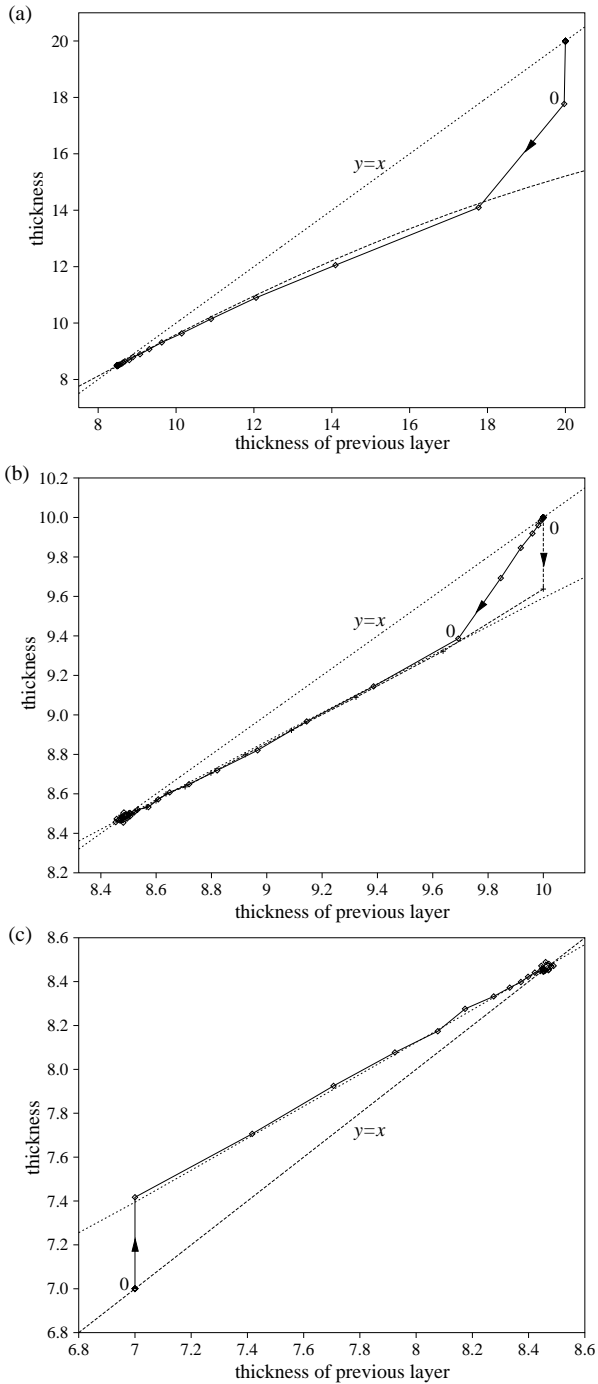,width=8.2cm}
\vglue 1mm
\begin{minipage}{8.5cm}
\caption{\label{fig:attract.square} 
Dependence of the thickness of a layer on the thickness of the previous layer.
The lines with data points in (a)--(c), ($l_{j-1},l_j$), are obtained from the crystal profiles in 
Figure \ref{fig:step.square}(a)--(c).
The arrows on these lines mark the direction of growth and the label `0' marks the point
at which $j$=0.
The dotted lines are $y=x$ and $l_n(l_{n-1})$, as labelled. $T=0.95\,T_m$.
}
\end{minipage}
\end{figure}
\end{center}

In Figure \ref{fig:step.square}b, as well as the average profile with respect to the 
position of the edge of the initial crystal, 
we also show the average profile measured with respect to the minimum position of the growth front
in each individual crystal. Now the sharp edge of the step is regained.
Therefore, the rounding of the step for the space-fixed profile can be understood to result from 
the variation in the amount by which the growth face transiently retreats back into the initial crystal.

To use the step profile to reveal the fixed-point attractor that describes the convergence to $l^{**}$ 
we plot in Figure \ref{fig:attract.square} the thickness of a layer against the 
thickness of the previous layer (i.e. the points ($l_{j-1},l_j$) where $j$ is the position of a layer) 
as we go through the step. 
The initial points from layers behind the step are all at ($l_{\rm init},l_{\rm init}$),
where $l_{\rm init}$ is the thickness 
of the initial crystal. As one passes through the step the points leave ($l_{\rm init},l_{\rm init}$)
and follow a path which converges to ($l^{**},l^{**}$).
It is the nature of this path that is of interest. 
In particular, the assumption of our previous suggestion that the steps produced by 
temperature changes could provide insight into the mechanisms of polymer crystallization was that 
this path should follow the fixed-point attractor, $l_n(l_{n-1})$.
$l_n(l_{n-1})$ is defined as the average thickness of a layer $n$ in the bulk of the crystal 
given that the thickness of the previous layer is $l_{n-1}$
and is obtained from the steady-state solution of the rate equations describing the SG model 
(for details see Ref. \onlinecite{Doye98e}).

In Figure \ref{fig:attract.square} we also plot $l_n(l_{n-1})$ to compare with the path taken by ($l_{j-1},l_j$).
For the crystal that is initially twenty units thick the path jumps from (20,20) to the fixed-point
attractor in two steps and then follows it down to the fixed point. 
The intermediate point is due to the slight rounding of the outer layer of the initial crystal.
The plot for the crystal that is initially 10 units thick is similar, except that the number of 
steps taken to reach the fixed-point attractor is larger because of the greater rounding of the 
the initial crystal. 
However, when the position is measured with respect to the minimum position of the growth front
the path of ($l_{j-1},l_j$) immediately steps onto the fixed-point attractor from the point (10,10).

\begin{center}
\begin{figure}
\epsfig{figure=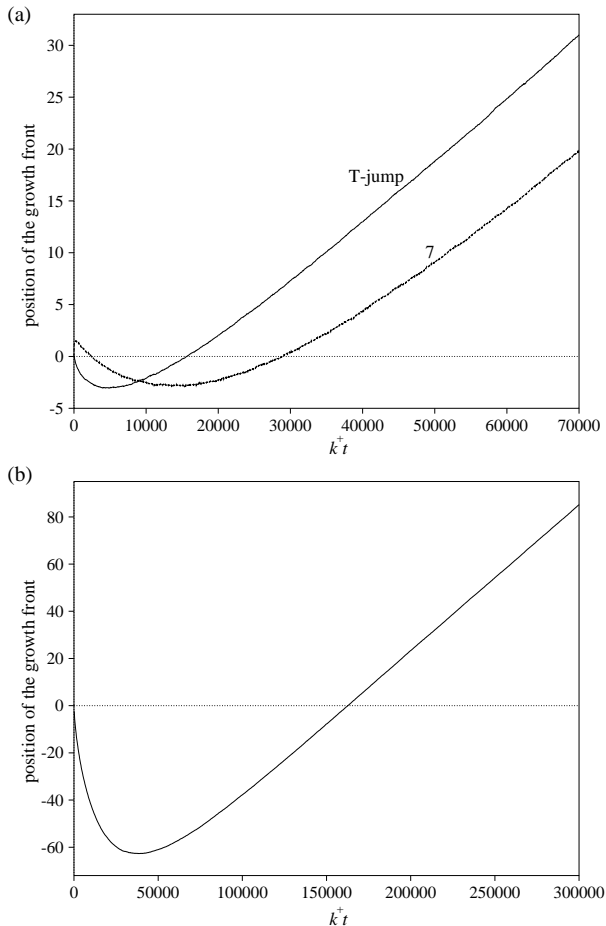,width=8.2cm}
\vglue 1mm
\begin{minipage}{8.5cm}
\caption{\label{fig:rate.thick} 
Position of the crystal growth front as a function of time.
In (a) the two lines are for growth from an initial crystal of thickness 7 
at $T=0.95\,T_m$ and for growth after a temperature jump from $0.935\,T_m$ to $0.95\,T_m$, as labelled.
(b) shows the growth after a temperature jump from $0.9\,T_m$ to $0.95\,T_m$.
}
\end{minipage}
\end{figure}
\end{center}

The third step profile in Figure \ref{fig:step.square} shows the profile when the thickness of the 
initial crystal is less than $l^{**}$. In this case the initial crystal is unstable with respect to
the melt/solution and so the growth face retreats (Figure \ref{fig:rate.thick}a). 
Only once a fluctuation to a greater thickness occurs does the growth face begin to advance.
Generation of this fluctuation is a slow process because there is an energetic cost associated
with the stems which overhang the previous layer. 
As the distance that the growth face initially goes backwards has a wide variation between individual
simulations, the gradient of the step in a space-fixed frame of reference is very shallow.
Only when the profile is measured with respect to the minimum position of the growth face 
does a clear picture of the thickening emerge. In this frame of reference there is a sharp 
upward step and the thickness quickly reaches $l^{**}$ (Figure \ref{fig:step.square}c). 
Furthermore, the path of ($l_{j-1},l_j$) jumps straight from (7,7) onto the thickening branch of the fixed-point 
attractor (Figure \ref{fig:attract.square}c). 

At this point it is right to consider which frame of reference---space-fixed or 
fixed with respect to the minimum position of the growth face---is more appropriate to the step profiles on real
polymer crystals. The profiles resulting from the two frames of reference represent two limits
in the degree of correlations between events along the growth face. 
The space-fixed frame of reference maintains the assumption of the two-dimensional SG model
that there are no correlations between adjacent stems along the growth face. 
It is this assumption that leads to the variation in the distance by which the growth face retreats.
However, this assumption is not always a good one. In particular, it seems likely that the nucleation of a 
thicker region would propagate laterally. Therefore, we expect the degree of correlation
to be closer to the limit obtained from using a frame of reference fixed with respect to the 
minimum position of the growth face. 
This latter approach 
is equivalent to assuming that the line of the step along the fold surface is straight, 
rather than rough.

Having shown that for the situation when growth occurs from a crystal that is of constant thickness 
a plot of ($l_{j-1},l_{j}$) can reveal the fixed-point attractor 
that underlies the growth mechanism of polymer crystals, we now proceed to consider
the steps that result from temperature jumps (from a temperature $T_1$ to a 
temperature $T_2$) during growth.
As the thickness of polymer crystals are approximately inversely proportional to the
degree of supercooling,\cite{Barham85} a decrease in temperature will lead to a downward step, 
and an increase in temperature will lead to an upward step.

\begin{center}
\begin{figure}
\epsfig{figure=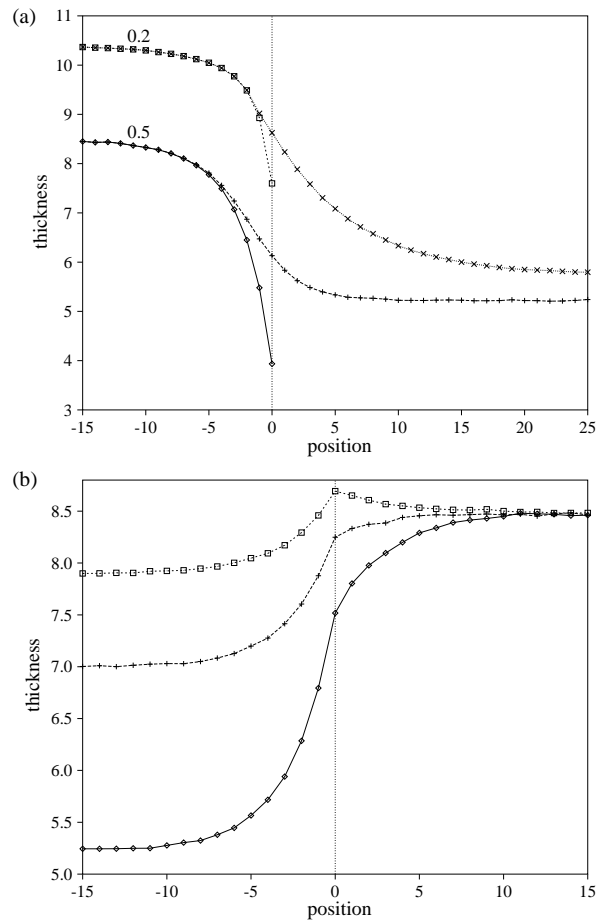,width=8.2cm}
\vglue 1mm
\begin{minipage}{8.5cm}
\caption{\label{fig:step.jump} 
Average crystal profiles that result from temperature jumps (a) from $0.95\,T_m$ to $0.9\,T_m$
with $kT_m/\epsilon$=0.5 and 0.2, as labelled, and (b) from $0.9\,T_m$ (solid line with diamonds),
$0.935\,T_m$ (dashed line with crosses) and $0.945\,T_m$ (dotted line with squares) to $0.95\,T_m$.
The zero of position is defined in (a) as the position of the
growth face when the temperature jump occurs
and in (b) as the minimum position of the growth face during the simulation. 
In (a) both the crystal profile before and after the temperature jump are
shown.
}
\end{minipage}
\end{figure}
\end{center}

\begin{center}
\begin{figure}
\epsfig{figure=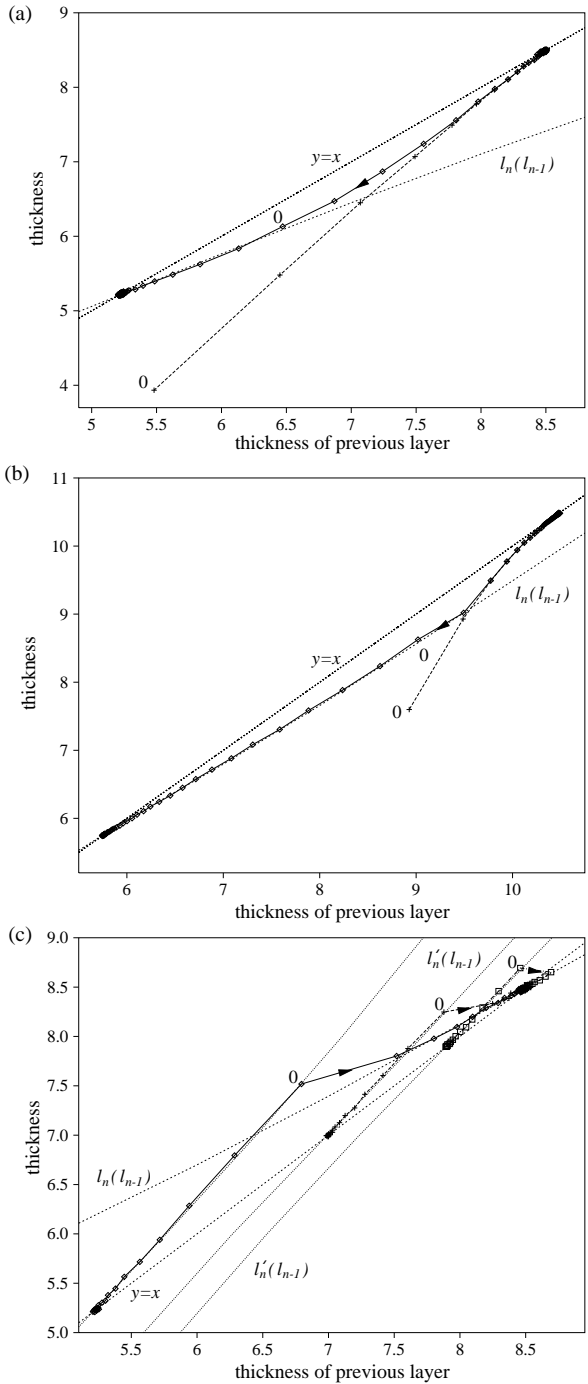,width=8.2cm}
\vglue 1mm
\begin{minipage}{8.5cm}
\caption{\label{fig:attract.jump} 
Dependence of the thickness of a layer on the thickness of the previous layer.
The lines with data points, ($l_{j-1},l_j$) are obtained from the step profiles in 
Figure \ref{fig:step.jump}.
The plots correspond to a temperature jump from $0.95\,T_m$ to $0.9\,T_m$
with (a) $kT_m/\epsilon$=0.5 and (b) $kT_m/\epsilon$=0.2 and (c)
to temperature jumps from $0.9\,T_m$ (solid line with diamonds), $0.935\,T_m$ (dashed line with crosses) and 
$0.945\,T_m$ (dotted line with squares) to $0.95\,T_m$. 
The arrows on these lines mark the direction of growth and the label `0' marks the point at which $j$=0.
In (a) and (b) the dashed line is for the profile of the crystal edge before the temperature
jump and the solid line for the step profile after the temperature jump.
The dotted lines are $y=x$, $l_n(l_{n-1},T_2)$ and $l'_n(l_{n-1},T_1)$, as labelled. 
}
\end{minipage}
\end{figure}
\end{center}

We consider the effects of a decrease in temperature first. In Figure 
\ref{fig:step.jump}a we show the average crystal profile before and after the decrease in
temperature. Firstly, unlike the situation considered above, the new growth after the
temperature change is on a crystal where there are variations in the layer thickness 
and where the layers near to the growth face are on average thinner. 
The rounded profile at the growing edge of the crystal is a characteristic property
of the SG model and plays a key role in Sadler and Gilmer's explanation of polymer
crystallization in terms of an entropic barrier.\cite{Sadler86a,Sadler87d,Sadler88a}

In the growth after the temperature jump not all of the rounding present at the 
edge of the crystal at the time of the jump is removed.
The profile of the resulting step initially follows the profile of the rounded edge, 
before changing curvature and smoothly converging to $l^{**}$ for the
new temperature (Figure \ref{fig:step.jump}a). 
The path of ($l_{j-1},l_{j}$) for the step initially leaves ($l^{**}(T_1),l^{**}(T_1)$)
and follows the same path as ($l_{j-1},l_{j}$) for the crystal edge at $T_1$. 
Only when this line meets $l_n(l_{n-1},T_2)$ does the path change slope and follow
the fixed-point attractor down to ($l^{**}(T_2),l^{**}(T_2)$) (Figure \ref{fig:attract.jump}a).

This basic scenario holds for all temperature decreases.
The main differences are only in the degree to which the step reflects the rounding of the
crystal edge, which in turn depends upon the relative slopes of $l_n(l_{n-1},T_2)$ and 
the path of ($l_{j-1},l_j$) for the crystal edge at $T_1$. 
For instance, if the product of the slopes is one then the crossover will occur halfway
between $l^{**}(T_1)$ and $l^{**}(T_2)$, and for the example shown in Figure \ref{fig:attract.jump}a
this is approximately the case.

The parameters in the model that can affect the two slopes are $kT_m/\epsilon$, $T_1$ and $T_2$.
We do not intend to survey the full parameter space, but instead just comment
on the effect of varying each parameter alone on the example in Figure 
\ref{fig:attract.jump}a.
For example, if we decrease $kT_m/\epsilon$ the slope of 
the fixed-point attractor becomes closer to one.
Therefore, the convergence of the thickness to $l^{**}(T_2)$ is more gradual (Figure 
\ref{fig:step.jump}a) and the path of ($l_{j-1},l_j$) follows the fixed point attractor to 
a greater extent (Figure \ref{fig:attract.jump}b).\cite{fold}
However, increasing $T_1$ has an opposite effect. It makes the slope of ($l_{j-1},l_j$)
for the crystal edge closer to one---the rounding is more gradual and extends deeper into the crystal
away from the crystal edge.  Therefore, the path of ($l_{j-1},l_j$) follows the fixed point attractor to
a lesser extent. Finally, changing $T_2$ has relatively little effect on the relative slopes and
so the crossover remains roughly midway between $l^{**}(T_1)$ and $l^{**}(T_2)$. 

In Figure \ref{fig:step.jump}b we show three examples of steps that result 
from temperature increases. In these cases a crystal of thickness of $l^{**}(T_1)$
is unstable with respect to the melt/solution at $T_2$ and so the crystal growth face
initially retreats after the temperature jump (Figure \ref{fig:rate.thick}).
In one of the cases ($T_1=0.935$) we chose $T_1$ so that $l^{**}(T_1)\approx 7$ 
enabling us to make a comparison with the step that was produced when the
initial crystal had a constant thickness of 7 units. From Figure \ref{fig:rate.thick}a
we can see that growth begins markedly earlier when there is a temperature jump.
The reason for this difference becomes clear when we examine the step 
profiles shown in Figure \ref{fig:step.jump}b. 
The majority of the step is behind the minimum position of the growth
face after the temperature jump. The growth face must retreat until it reaches a region
of the crystal where the layer thickness is larger than the average value at $T_1$. 
New growth then begins from this position and the crystal thickens the small
amount necessary to reach $l^{**}(T_2)$. Only for this new growth does ($l_{j-1},l_j$)
follow the fixed-point attractor (Figure \ref{fig:attract.jump}c).
Growth is more rapid than for the case where growth is from an initial crystal of constant thickness 
because encountering already present fluctuations during the retreat of the growth
face is a more common event than the generation of new fluctuations to larger thickness.

Interestingly, the part of the step resulting from already present fluctuations
has a well-defined behaviour. In the bulk of the crystal (where the influence of the
growth face can no longer be felt) there is a symmetry between the directions towards
and away from the growth face. Therefore, $l_{n-1}(l_n)=l_{n}(l_{n-1})$; i.e.\ in
the bulk of the crystal the 
dependence of the thickness of layer $n-1$ on the thickness of layer $n$ is the same
as the dependence of the thickness of layer $n$ on the thickness of layer $n-1$.
Given that at the minimum position of the growth face there is 
a fluctuation to a certain amount larger than $l^{**}$, 
the layers behind this would therefore be expected to obey $l_{n-1}(l_n,T_1)$.
So, it is unsurprising that a plot of $(l_j,l_{j-1})$ for $j\le 0$ 
follows the fixed point attractor for $T_1$.
From this it simply follows that a plot of $(l_{j-1},l_j)$ for $j\le 0$ follows
the curve, $l'_n(l_{n-1},T_1)$, formed by reflecting $l_n(l_{n-1},T_1)$ in $y=x$ (Figure 
\ref{fig:attract.jump}c).
For $j>0$ 
$(l_{j-1},l_j)$ jumps from $l'_n(l_{n-1},T_1)$ onto the fixed point attractor for $T_2$.

From Figures \ref{fig:rate.thick}, \ref{fig:step.jump}b and \ref{fig:attract.jump}c
we can ascertain some of the effects of changing the size of the temperature increase.
When the temperature jump is larger the growth face retreats further because a
larger fluctuation in the thickness needs to be encountered to nucleate new growth.
Furthermore, the larger temperature jump the more of the step is a result of new growth.
Interestingly, for the smallest temperature increase the growth face retreats to a fluctuation
in thickness which is on average larger than $l^{**}(T_2)$.
Therefore, the profile displays a lip (Figure \ref{fig:step.jump}b)
and in the new growth the crystal thins slightly to reach the new $l^{**}$.

The effect of $kT_m/\epsilon$ is somewhat similar to the
effect of the magnitude of the temperature increase.
A larger proportion of the step is the result of new growth
when $kT_m/\epsilon$ is smaller.
Presumably, this is because there is less variation in the stem length 
for smaller values of $kT_m/\epsilon$.\cite{Spinner95} 

\section{Discussion}

In this paper we have seen that the step profiles obtained by simulations of temperature 
jumps using the SG model do reveal information about the fixed-point attractor that
we have recently argued underlies the mechanism of polymer crystallization.\cite{Doye98b,Doye98d,Doye98e}
This 
strengthens our suggestion that temperature jump experiments 
can provide
insight into the physics of polymer crystallization.
However, 
the step profiles for temperature decreases also reflect
the rounding of the crystal edge and those for temperature increases also reflect the variations in
the stem length that are present in the crystals.
These additional features mean that the interpretation of experimental
step profiles in terms of fixed-point attractor curves requires considerable care.
This disadvantage is partly offset by the the fact that an experimental study of the step profiles 
could reveal information about aspects of polymer crystallization not originally anticipated.

However, we note that it is not clear to what extent these additional features occur for
the steps on real polymer crystals---they may just
reflect some of the simplifications in the SG model. 
Firstly, although the crystal edge is always rounded in the SG model, there is,
as far as we are aware, not yet any direct experimental evidence of this effect 
occurring in real polymer crystals under normal crystallization conditions. 
Secondly, in an alternative model of polymer crystallization, rounding 
only occurs at small supercoolings.\cite{Doye98b,Doye98d}

Secondly, it may be that the SG model overestimates the roughness of the
fold surface of polymer crystals. 
In the SG model changes in the length of a stem can occur only when that stem is 
at the growth face. Once the growth face has passed through a region the stem length
variations that arise from the kinetics of growth are frozen in. However, it may be that
in real crystals there are annealing mechanisms (which only require local motion of stems)
that can act to reduce the magnitude of these fluctuations to their equilibrium values. 
Such processes might make the crystal growth following a temperature increase more similar
to the growth from an initial crystal of constant thickness less than $l^{**}$.\cite{SGanneal}
Recent atomic-force microscopy experiments on the fold surface of polyethylene\cite{Patil94,Magonov97}
shed some light on this issue. Only for a minority of crystal could the folds
be clearly resolved. It was suggested that the lack of clear order for the majority of
crystals resulted from differences in the heights of the folds, i.e.\ there is some surface roughness.

\acknowledgements
The work of the FOM Institute is part of the research programme of
`Stichting Fundamenteel Onderzoek der Materie' (FOM)
and is supported by NWO (`Nederlandse Organisatie voor Wetenschappelijk Onderzoek'). 
JPKD acknowledges the financial support provided by the Computational Materials Science 
program of the NWO and by Emmanuel College, Cambridge.

\end{multicols}

\begin{thebibliography}{10}
\bibitem{GGWstep}
In Ref.\ \onlinecite{Goldbeck90a} there is a brief presentation of results that
  confirm that in the SG model temperature jumps do give rise to steps.
\bibitem{GGWrate}
An extended version of the rate equation approach can also be used to study the
  transient behaviour of the SG model.\cite{Goldbeck89a,Goldbeck93a} However,
  the frame of reference used (fixed with respect to the growth face) is not
  the most convenient for the purposes of this paper.
\bibitem{fold}
If we introduce an energetic contribution from the fold surface (as in Ref.\
  \onlinecite{Doye98e}) a similar effect to that which results from decreasing
  $kT_m/\epsilon$ is observed.
\bibitem{SGanneal}
If such annealing mechanisms were to occur they might also be able to change
  the shape of the steps after they have formed, perhaps scrambling some of the
  information on the crystallization mechanism that they contain.
\end{thebibliography}
\end{document}